\documentclass[12pt]{article}%
\usepackage{color}
\usepackage{fullpage}
\usepackage{amsmath}%
\usepackage{amssymb}%
\usepackage{amscd}%
\usepackage{amsthm}
\usepackage{graphicx}
\usepackage{amssymb}
\usepackage{epstopdf}
\usepackage{cancel}
\usepackage{setspace}
\usepackage{slashed}
\usepackage[hang,small]{caption}
\def\p{\partial}

\def\R{\mbb{R}}

\def\half{\frac{1}{2}}

\newcommand{\be}{\begin{equation}}
\newcommand{\ee}{\end{equation}}
\newcommand{\bra}[1]{\langle #1|}
\newcommand{\ket}[1]{|#1\rangle}
\newcommand{\braket}[2]{\langle #1|#2\rangle}
\newcommand{\mbb}[1]{\mathbb{#1}}

\newcommand{\pd}[2]{\frac{\p#1}{\p#2}}

\theoremstyle{definition}

\numberwithin{equation}{section}
\begin{document}
%---Title ---------------------------------------------------------------
\begin{titlepage}
\bigskip
\rightline{}

\bigskip\bigskip\bigskip\bigskip
\centerline {\Large \bf {Lifshitz Singularities}}
\bigskip\bigskip

\centerline{\large  Gary T. Horowitz and Benson Way}
\bigskip\bigskip
\bigskip\bigskip
\centerline{\em Department of Physics, UCSB, Santa Barbara, CA 93106}
\centerline{\em  gary@physics.ucsb.edu, benson@physics.ucsb.edu}
\bigskip\bigskip
\begin{abstract}Lifshitz spacetimes are possible gravitational duals to strongly coupled field theories with an anisotropic scaling symmetry.  These spacetimes however, have a null curvature singularity.  We find that higher dimensional embeddings of Lifshitz also have a similar singularity.  We study the propagation of test strings in this background and find that they become infinitely excited if they try to propagate through the singularity.  This means that the Lifshitz geometry is unstable and will receive large corrections in string theory.
\end{abstract}
\end{titlepage}

%\tableofcontents

%===MAIN=================================================================
\onehalfspacing
\begin{section}{Introduction}

The AdS/CFT correspondence \cite{Maldacena98,Gubser:1998bc,Witten:1998qj} provides weakly coupled and calculable gravitational descriptions of certain strongly coupled field theories.  This is a realization of holography -- the idea that a non-gravitational theory is equivalent to a gravitational theory in a higher dimension.  While the original AdS/CFT correspondence describes a duality between a conformal field theory and a string theory, the idea of holography seems to be much broader.  The more general gauge/gravity duality has been explored as a means to describe a wide range of strongly coupled systems within QCD \cite{Kovtun:2004de} and condensed matter physics \cite{Hartnoll:2009sz,McGreevy:2009xe}.  

In particular, a gravitational description of Lifshitz-like fixed points was proposed \cite{Kachru:2008yh}.  In the context of condensed matter, various systems exhibit a dynamical scaling near fixed points:
\be
t\rightarrow\lambda^zt\;,\qquad x\rightarrow\lambda x,\qquad z\neq1\;.
\label{scaling}
\ee
That is, rather than obeying the conformal scale invariance,
$t\rightarrow\lambda t, \ x\rightarrow\lambda x,$
 the temporal and spatial coordinates scale anisotropically.  Also imposing invariance under time and space translations, spatial rotations, spatial parity, and time reversal, the authors of \cite{Kachru:2008yh} propose the following $D$-dimensional spacetime metric:
 \be
ds^2=\ell^2\left(-r^{2z}dt^2+\frac{dr^2}{r^2}+r^2 d x_idx^i\right)\;,
\label{Lifshitz}
\ee
where $2\leq i \leq D-1$. These spacetimes obey the scaling relation \eqref{scaling} if one also scales $r\rightarrow\lambda^{-1}r$.  If $z=1$, this spacetime is the usual AdS metric in Poincar\'{e} coordinates with AdS radius $\ell$.  

Metrics of the form $\eqref{Lifshitz}$ can be obtained as solutions to general relativity with a negative cosmological constant and appropriate matter content.  For example, solutions were found by introducing one and two-form gauge fields \cite{Kachru:2008yh}, a massive vector field \cite{Taylor:2008tg} (or abelian Higgs model \cite{Gubser:2009cg}), or a charged perfect fluid \cite{Hartnoll:2010gu}.  Black hole solutions with Lifshitz asymptotics were also found   \cite{Mann:2009yx,Bertoldi:2009vn,Balasubramanian:2009rx,Danielsson:2009gi}.  There are by now many embeddings of Lifshitz in supergravity and string theory following the original work of  \cite{Hartnoll:2009ns,Balasubramanian:2010uk,Donos:2010tu,Gregory:2010gx}.  

However, there is a problem with the Lifshitz metric (\ref{Lifshitz})  as $r \rightarrow 0$.  Despite the fact that all scalar curvature invariants are constant, there is a curvature singularity if $z\neq1$ \cite{Hartnoll:2009sz,Kachru:2008yh,Copsey:2010ya}.  This can be seen by computing the tidal forces between infalling geodesics.

We study the nature of this singularity in string theory. The starting point is the observation of Adams et al. \cite{Adams:2008zk}  that 
perturbative stringy corrections only renormalize $z$ and the AdS radius $\ell$. So the Lifshitz spacetime is a solution to all orders in $\alpha'$ and these corrections cannot resolve the singularity unless $z$ is driven to one.  One might hope that the higher dimensional embeddings of Lifshitz will be free of singularities, but we will show that this is not the case.

Of course, not all spacetimes that are singular in the sense of general relativity (i.e. geodesically incomplete) are singular in the sense of string theory (e.g. orbifold spacetimes).  To see if a spacetime is singular in string theory we must study the motion of test strings. It turns out that the Lifshitz singularity is identical to a singular plane wave. In retrospect this is not surprising since the singularity is null and the spacetime is homogeneous in the $D-2$ transverse directions. Fortunately, string propagation in plane-wave backgrounds was studied in the early 1990's  \cite{Horowitz:1989bv,deVega:1990ke}. Using those results we will see that  strings trying to propagate through a Lifshitz singularity  become infinitely excited, and hence  the Lifshitz spacetime is indeed singular in string theory.  This means that quantum corrections are important and the Lifshitz metric does not describe the far infrared physics of a Lifshitz critical point.

In the following section, we review the tidal forces in the Lifshitz geometry and show that the higher dimensional embeddings are also singular.  In section 3, we motivate the plane-wave approximation to Lifshitz spacetimes near the singularity, and in the following section we analyze the motion of test strings in this plane-wave background. The final section has some concluding comments.  
\end{section}

\begin{section}{Tidal Forces}

We start by considering tidal forces in the Lifshitz metric \eqref{Lifshitz}.   In these coordinates, the components of the Riemann tensor are finite, and therefore, all curvature invariants constructed from the Riemann tensor are also finite.  Nevertheless,  if $z\neq 1$ there is a curvature singularity at $r=0$ due to diverging tidal forces \cite{Hartnoll:2009sz,Kachru:2008yh,Copsey:2010ya}.  This singularity is also reached in finite proper time by infalling observers so the spacetime is geodesically incomplete.  For $z=1$, the metric is the familiar AdS metric in Poincar\'{e} coordinates, and the would-be curvature singularity is merely a coordinate singularity.  

Consider a radial timelike geodesic with tangent vector $T=(\dot t,\dot r,\vec0)$, where the dots denote $d/d\tau$.  There is a conserved energy $E=\dot t \ell^2r^{2z}$, and the normalization $T_\mu T^\mu=-1$ gives
\be
\dot r^2=\frac{E^2r^{2(1-z)}}{\ell^4}\left(1-\frac{\ell^2r^{2z}}{E^2}\right)\;.
\ee
Now we choose an orthonormal frame parallelly propagated along such a geodesic:
\begin{subequations}
\begin{align}
(e_0)^\mu&=\frac{E}{\ell^2 r^{2z}}\left(\pd{}{t}\right)^\mu-\frac{Er^{1-z}}{\ell^2}\sqrt{1-\frac{\ell^2r^{2z}}{E^2}}\left(\pd{}{r}\right)^\mu\;,\\
(e_1)^\mu&=\frac{E}{\ell^2 r^{2z}}\sqrt{1-\frac{\ell^2r^{2z}}{E^2}}\left(\pd{}{t}\right)^\mu-\frac{Er^{1-z}}{\ell^2}\left(\pd{}{r}\right)^\mu\;,\\
(e_i)^\mu&=\frac{1}{\ell r}\left(\pd{}{x^i}\right)^\mu\;.
\end{align}
\end{subequations}
Then using the notation 
\be
R_{abcd}=R_{\mu\nu\rho\sigma}(e_a)^\mu(e_b)^\nu(e_c)^\rho(e_d)^\sigma,
\ee
the nonzero components of the Riemann tensor in this frame are given by (no sum over repeated indices)
\begin{subequations}
\begin{align}
R_{0101}&=\frac{z}{\ell^2}\;,\\
R_{ijij}&=-\frac{1}{\ell^2}\qquad (i\neq j)\;,\\
R_{0i0i}&=\frac{E^2(z-1)}{\ell^4r^{2z}} +\frac{1}{\ell^2}\;,\\
R_{1i1i}&=\frac{E^2(z-1)}{\ell^4r^{2z}} -\frac{z}{\ell^2}\;,\\
R_{0i1i}&=\frac{E^2(z-1)}{\ell^4 r^{2z}}\sqrt{1-\frac{\ell^2r^{2z}}{E^2}}\;.
\end{align}
\label{Lifshitztidal}
\end{subequations}
Thus, if $z\neq 1$ (and $z>0$), the tidal forces diverge as $(z-1)/{r^{2z}}$.  

Since constant $r$ surfaces are timelike, their limit as $r\rightarrow 0$  must be either timelike or null. The vectors normal to surfaces of constant $r$ become null as $r\rightarrow 0$: 
\be
\nabla_\mu r\nabla^\mu r= g^{rr}=\frac{r^2}{\ell^2}
\ee
 suggesting that $r=0$ is a null curvature singularity. A more precise way to show this is to consider radial null geodesics. If the singularity were timelike, an outgoing light ray from $r=0$ could lie entirely to the future of an ingoing one. However radial null geodesics satisfy $dt = \pm dr/r^{1+z}$, so $t\rightarrow \pm \infty$ as $r\rightarrow 0$ showing this is impossible. 

Let us now comment on how these singularities arise when the Lifshitz metric  comes from a higher dimensional spacetime. There are two broad classes of embeddings in supergravity or string theory. We now demonstrate that both classes suffer from singularities due to diverging tidal forces.  

In the first approach, the metric takes the form  \cite{Balasubramanian:2010uk,Donos:2010tu,Cassani:2011sv,Halmagyi:2011xh,Narayan:2011az,Chemissany:2011mb} 
\be
ds^2=r^2(2d \sigma dt+dx_idx^i)+\gamma \frac{dr^2}{r^2}+f\,d\sigma^2+ds_E^2\;,
\ee
where $\gamma$ is a dimension dependent constant, $ds^2_E$ is the metric on some Sasaki-Einstein manifold, and $f$ is a function of $\sigma$ and the coordinates on $E$. This metric looks like AdS with an extra line and Sasaki-Einstein manifold added, and appears to be nonsingular. However, it can be rewritten in the form
\be
ds^2 = \left[ -\frac{ r^4}{f} dt^2 + r^2 dx_idx^i+\gamma \frac{dr^2}{r^2}\right]+f(d\sigma + \frac{r^2}{f} dt)^2 +ds_E^2
\ee
If $\sigma$ is periodic and $f$ is constant, one can do a standard Kaluza-Klein reduction and  obtain the
 Lifshitz metric with $z=2$. Even when $f$ is not constant, one can argue that the effective geometry on scales large compared to the compact directions will look like Lifshitz. One does not usually create singularities by dimensional reduction on a circle unless that circle becomes null (or pinches off) \cite{Gibbons:1994vm}, which is not the case here. So one expects that the original metric must itself be singular.\footnote{This was also shown in \cite{Chemissany:2011mb}.}
 
To establish this,
it suffices to show that one component of the Riemann curvature tensor diverges in a parallelly propagated orthonormal frame.  We will be concerned with the component $R_{0i0i}$.  Consider the tangent vector  $T=(\dot t,\dot\sigma,\dot r,\vec 0)$.  The Killing field $\partial/ \partial t$ gives a conserved energy $E=\dot\sigma r^2$.  Then the normalization $T_\mu T^\mu=-1$ implies
\be
\dot t=-\frac{1}{2E}\left(1+\frac{E^2f}{r^4}+\gamma\, \frac{\dot r^2}{r^2}\right)\;.
\ee
 In order for the vector $T^\mu$ to be tangent  to a geodesic,  $r(\tau)$ must  solve the geodesic equation. But even without knowing the solution explicitly, one can show there is a singularity as follows.  Two basis vectors of an orthonormal frame parallelly propagated along this geodesic are
\begin{subequations}
\begin{align}
(e_0)^\mu&=-\frac{1}{2E}\left(1+\frac{E^2f}{r^4}+\gamma\, \frac{\dot r^2}{r^2}\right)\left(\pd{}{t}\right)^\mu+\frac{E}{r^2}\left(\pd{}{\sigma}\right)^\mu+\dot r\left(\pd{}{r}\right)^\mu\;,\\
(e_i)^\mu&=\frac{1}{r}\left(\pd{}{x^i}\right)^\mu\;.
\end{align}
\end{subequations}
It follows that
\be
R_{0i0i}=\frac{1}{\gamma}\left(1+\frac{E^2f}{r^4}\right)\;.
\ee
Note that this component does not depend on $\dot r$.  Comparing with \eqref{Lifshitztidal}, we see that this component of the tidal forces diverges in a similar way to Lifshitz with $z=2$.  

A related construction in  \cite{Singh:2010zs} yields Lifshitz with $z=3$. A similar argument shows that the higher dimensional solution again has a curvature singularity.

The other class of higher dimensional solutions consist of a warped product of Lifshitz with some other space \cite{Gregory:2010gx,Donos:2010ax}.  They are schematically of the form
\be
ds^2=f(\rho)ds_{Li}^2+g(\rho)d\rho^2+ds_\rho^2\;.
\ee
where $ds_{Li}^2$ is the Lifshitz metric (\ref{Lifshitz}), $ds_\rho^2$ is the metric for some space with possible dependence on the $\rho$ coordinate.  (For simplicity, we will absorb the AdS radius $\ell$ into the function $f$.)  Let us fix the coordinate $\rho=\rho_0$ and choose a tangent vector $T=(\dot t,\dot r,\vec 0)$.  The conserved quantity $E$ and  the usual normalization $T_\mu T^\mu=-1$ lets us write down two of the components of an orthonormal frame
\begin{subequations}
\begin{align}
(e_0)^\mu&=\frac{E}{r^{2z}f(\rho_0)}\left(\pd{}{t}\right)^\mu-\frac{Er^{1-z}}{f(\rho_0)}\sqrt{1-\frac{r^{2z}f(\rho_0)}{E^2}}\left(\pd{}{r}\right)^\mu\;,\\
(e_i)^\mu&=\frac{1}{r\sqrt{f(\rho_0)}}\left(\pd{}{x^i}\right)^\mu\;.
\end{align}
\end{subequations}
In general, $T^\mu$ is not  tangent  to a geodesic.  However, observers can follow a path with tangent vector $T^\mu$ with a constant acceleration.  The norm of the acceleration $A^\mu=T^\nu\nabla_\nu T^\mu$ is
\be
A^\mu A_\mu=\frac{f'(\rho_0)^2}{4f(\rho_0)^2g(\rho_0)}\;.
\ee
Note that this is independent of $E$ and $r$.  As long as  $f(\rho_0)$  and $g(\rho_0)$ are nonzero, this acceleration is finite, and if $f'(\rho_0)=0$  this curve is a geodesic.  Even when the curve is not a geodesic,  computing the Riemann curvature in this frame gives
\be
R_{0i0i}=\frac{E^2(z-1)}{f(\rho_0)^2r^{2z}} + \frac{f'(\rho_0)^2}{4f(\rho_0)^2g(\rho_0)}+\frac{1}{f(\rho_0)}\;.
\ee
Therefore, up to a factors of $f(\rho_0)$, these warped products of Lifshitz also suffer from the same singularities.  
\end{section}

\begin{section}{A Plane Wave Approximation}

As mentioned in the introduction, the fact that the Lifshitz singularity is null and the spacetime is homogeneous in the transverse directions suggests that the region near the singularity can be modeled by a plane wave\footnote{A similar approximation was done in a  different context in \cite{Horowitz:1997ed}.}. We now demonstrate this explicitly.

 First, let us define the tortoise coordinate $r_*$ such that
\be
dr_*=r^{-1-z}dr,\qquad r_*=-\frac{1}{z r^z}\;,
\ee
and then define the null coordinates $u=t-r_*$, $v=t+r_*$.  The metric (\ref{Lifshitz}) becomes
\begin{subequations}
\begin{align}
ds^2&=\frac{\ell^2}{z^2r_*^2}(-dt^2+dr_*^2)+\ell^2\left(\frac{1}{z^2r_*^2}\right)^{1/z}d x_idx^i\;,\\
&=-\frac{4\ell^2}{z^2(u-v)^2}dudv+\ell^2\left(\frac{4}{z^2(u-v)^2}\right)^{1/z}d x_idx^i\;.
\end{align}
\end{subequations}
From the coordinate transformations, we see that
\be
r^z=-\frac{1}{zr_*}=\frac{2}{z(u-v)},
\ee
so small $r$ corresponds to $u\gg v$.  Then near the singularity $r=0$, we can make the approximation
\be\label{approx}
ds^2\approx-\frac{4\ell^2}{z^2u^2}dudv+\ell^2\left(\frac{4}{z^2u^2}\right)^{1/z}d x_idx^i\;.
\ee
Now let $u=-{4\ell^2}/{z^2U}$.  From this coordinate transformation, small $r$ is approximated by
\be
r^{z}\approx -\frac{zU}{2\ell^2}\;,
\label{rtou}
\ee
so we should study string propagation in this metric to $U=0$ from $U<0$.  Our line element becomes
\be
ds^2\approx-dUdv+\ell^2\left(\frac{zU}{2\ell^2}\right)^{2/z}d x_idx^i\;.
\ee
This is a plane wave metric.  To bring it into the form used in \cite{deVega:1990ke} we use the change of coordinates 
\be
v=V-\frac{1}{zU}X_iX^i,\qquad x_i=\frac{X_i}{\ell}\left(\frac{zU}{2\ell^2}\right)^{-1/z}\;,
\ee
Then the metric becomes
\be
ds^2\approx-dUdV+dX_idX^i+W(U)X_iX^idU^2\;,\qquad W(U)=\frac{1-z}{z^2U^2}\;.
\label{pwave}
\ee

If $z=1$, the Lifshitz metric (\ref{Lifshitz}) is AdS and (\ref{pwave}) is the metric for Minkowski space.   Since $r=0$ (the Poincar\'{e} horizon) is merely a coordinate singularity, test strings will have no trouble crossing it.  Similarly, $U=0$ causes no trouble for strings in Minkowski space.  But even though (\ref{pwave}) captures this property of (\ref{Lifshitz}),  we have replaced a spacetime with a cosmological constant with one that is Ricci flat. In other words, the approximation (\ref{approx}) removes the cosmological constant, so the plane wave metric does not adequately describe Lifshitz when $z=1$.  

However if $z\neq1$, the dynamics of test strings close to the curvature singularity $r=0$ are dominated by the diverging tidal forces.  As we will now show, the tidal forces near $U=0$ for the metric (\ref{pwave}) behave in exactly the same way.  Therefore, the Lifshitz metric (\ref{Lifshitz}) near the singularity is well-approximated by this plane wave.  

As in the previous section, we consider radial timelike geodesics with a tangent vector $T=(\dot U,\dot V,\vec0)$.  The killing vector $\partial/\partial V$ gives a conserved energy $E=\half\dot U$, and the normalization $T_\mu T^\mu=-1$ gives
\be
\dot V=\frac{1}{2E}\left(4E^2W(U)X_iX^i+1\right)\;.
\ee
Now we choose the parallelly propagated orthonormal frame
\begin{subequations}
\begin{align}
(e_0)^\mu&=2E\left(\pd{}{U}\right)^\mu+\frac{1}{2E}\left(4E^2W(U)X_iX^i+1\right)\left(\pd{}{V}\right)^\mu\;,\\
(e_1)^\mu&=2E\left(\pd{}{U}\right)^\mu+\frac{1}{2E}\left(4E^2W(U)X_iX^i-1\right)\left(\pd{}{V}\right)^\mu\;,\\
(e_i)^\mu&=\left(\pd{}{X^i}\right)^\mu\;.
\end{align}
\end{subequations}
The nonzero components of the Riemann tensor in this frame are given by
\be
R_{0i0i}=R_{1i1i}=R_{0i1i}=-4E^2W(U)=\frac{4E^2(z-1)}{z^2U^2}\;.
\ee
Using (\ref{rtou}), and comparing this with (\ref{Lifshitztidal}), we see that both metrics have diverging tidal forces that act in the same directions and diverge as $(z-1)/{r^{2z}}$.  We therefore conclude that the behavior of test strings near the null singularity of Lifshitz can be well approximated by the plane wave metric (\ref{pwave}).

\end{section}

\begin{section}{Test Strings}

We now study the behavior of (first quantized)  test strings in the plane-wave metric  (\ref{pwave}).  This is essentially identical to a calculation that was done in \cite{deVega:1990ke}. The only difference is that \cite{deVega:1990ke} considers vacuum solutions so the $X^i X_i$ factor in (\ref{pwave}) is replaced by $X^2 - Y^2$. For completeness, we review the calculation below.

  The motion of strings on a given background is described by the action
\be
S=-\frac{1}{4\pi\alpha'}\int\;d\tau d\sigma\sqrt h \;h^{ab}g_{\mu\nu}(X)\partial_a X^\mu\partial_b X^\nu\;,
\ee
where $X^\mu=X^\mu(\sigma,\tau)$ is the embedding of the string world sheet in spacetime, $h_{ab}$ is the world-sheet metric, and $\alpha'$ is the inverse string tension.  Weyl invariance and reparametrization invariance allows us to choose the conformal gauge $h_{ab}=e^{\phi(\sigma,\tau)}\eta_{ab}$.  Since the metric is a plane wave, we also can work in light-cone gauge $U=\alpha'p\tau$ \cite{Horowitz:1989bv}.  If we decompose the $X^i$ into modes
\be
X^i(\sigma,\tau)=\sum_{n=\infty}^\infty X^i_n(\tau)e^{in\sigma}\;,
\ee
the worldsheet equations of motion for $X^i$ become
\be
\ddot X_n^i+\left(n^2-{\alpha'}^2p^2W(\alpha' p\tau)\right)X^i_n=0\;,
\label{Xeom}
\ee
where the dot denotes differentiation by $\tau$.  This equation is just like a one-dimensional Schr\"{o}dinger equation for a particle of energy $n^2$ in a potential $\alpha'^2p^2W$. Dividing \eqref{Xeom} by $n^2$ shows that the modes must be functions of $n\tau$:
\be
X_n^i(\tau)=X_1^i(n\tau)\;.
\label{ntau}
\ee  
The component $V(\sigma,\tau)$ is determined by
\be
\alpha'p\dot V=\dot X_i^2+{X_i'}^2+{\alpha'}^2p^2 W(\alpha' p\tau)X_iX^i\;,\qquad \alpha'pV'=2\dot X_i X_i'\;.
\ee

To make it easy to identify the excited state of the string, we will consider a plane wave with flat spacetime regions before and after it.  Accordingly, we pick a large time $T$, set  $W(U)=0$ for $|U|\geq T$ and choose $W(U)$ to reproduce the Lifshitz singularity for $|U|\leq T$.  Thus
\be
\begin{split}
W(U)&=\pm k\left( \frac{1}{U^2}-\frac{1}{T^2}\right)\;,\qquad |U|\leq T\;,\\
W(U)&=0\;,\qquad |U|\geq T\;,
\end{split}
\label{newW}
\ee
where without loss of generality, we have chosen $k>0$.  Given a choice of $z$, the sign and the value of $k$ can be determined according to (\ref{pwave}): 
\be
\pm k=\frac{1-z}{z^2},\qquad k>0\;.
\label{keq}
\ee
In particular, $z>1$ implies $W<0$ and an attractive potential, while $z<1$ implies $W>0$ and a repulsive potential.  

With this choice of $W$, the $X^i$ are given by the usual flat space expansions in the region $U\leq-T$:
\be
\begin{split}
X^i(\sigma,\tau)&=q_<^i+2\alpha'p_<^i\tau+\sum_{n\neq0}e^{in\sigma}X_n^i(\tau)\;,\qquad \tau\leq-\tau_0\;,\\
X_n^i(\tau)&=i\frac{\sqrt{\alpha'}}{n}\left(\alpha^i_{n<}e^{-in\tau}-\tilde\alpha^i_{-n<}e^{in\tau}\right)\;,\qquad n\neq0\;,
\end{split}
\label{flatsol}
\ee
where $\tau_0=T/\alpha'p$ and the mode operators $\alpha^i_{n<},$ and $\tilde\alpha^i_{n<}$ satisfy the usual canonical commutation relations.  

In the region $|U|\leq T$, the equations of motion for $X^i$ can be solved in terms of Bessel functions \cite{deVega:1990ke}.  For our purposes, it suffices to examine the solutions near the singularity $U\rightarrow0$.  In that case, the $X_n^i$ satisfy
\be
\ddot X_n^i\mp \frac{k}{\tau^2}X_n^i=0\;.
\ee
This can be solved exactly, so the solutions for $\tau\rightarrow0^-$ behave as
\be
X_n^i(\tau)=C_n^i\left|n\tau\right|^{\tfrac{1}{2}(1-\nu_\pm)}+D_n^i\left|n\tau\right|^{\tfrac{1}{2}(1+\nu_\pm)}\;,
\label{x0sol}
\ee
where $\nu_{\pm}=\sqrt{1\pm4k}$\;.  

If $z<1$, the positive sign is chosen in \eqref{newW} and \eqref{keq}, and $\nu_+$ must be used in \eqref{x0sol}.  Since $\nu_+>1$ and a generic solution has $C^i_n\neq0$, $X^i$ will tend to infinity when the string approaches the singularity $\tau\rightarrow0$; the repulsive potential pushes the string away in the transverse directions.  Therefore, a generic string will not pass through the singularity and instead becomes infinitely large in a finite time $\tau$. 

If instead $z>1$, the negative sign is chosen and $\nu_-$ is used.  The attractive potential pulls  $X^i$ towards the origin.  Eventually, the string will hit the singularity in finite time.  From \eqref{keq}, we see that $\nu_-$ remains real so the solutions do not oscillate near the singularity.  

In the region $U\geq T$, the solutions to $X^i$ are again given by the expansion in flat spacetime: 
\be
\begin{split}
X^i(\sigma,\tau)&=q_>^i+2\alpha'p_>^i\tau+\sum_{n\neq0}e^{in\sigma}X_n^i(\tau)\;,\qquad \tau\geq\tau_0\;,\\
X_n^i(\tau)&=i\frac{\sqrt{\alpha'}}{n}\left(\alpha^i_{n>}e^{-in\tau}-\tilde\alpha^i_{-n>}e^{in\tau}\right)\;,\qquad n\neq0\;.
\label{flatsol2}
\end{split}
\ee
The operators $\alpha^i_{n>}$, $\tilde\alpha^i_{n>}$ are related to those in \eqref{flatsol} $\alpha^i_{n<}$, $\tilde\alpha^i_{n<}$ by the Bogoliubov transformation
\be
\begin{split}
\alpha^i_{n>}=A_n^i\alpha^i_{n<}+B_n^i\tilde\alpha^{i\dagger}_{n<}\;,\\
\tilde\alpha^i_{n>}=A_n^i\tilde\alpha^i_{n<}+B_n^i\alpha^{i\dagger}_{n<}\;.
\label{Bog}
\end{split}
\ee

The solutions of \eqref{Xeom} with the boundary condition
\be
f_n^i(\tau)=e^{in\tau},\qquad \tau<-\tau_0
\ee
can be written in the following implicit integral form:
\be
f_n^i(\tau)=e^{in\tau}+\frac{p^2\alpha'^2}{2in}\left(e^{in\tau}\int_{-\infty}^{\tau}d\tau'\;e^{-in\tau'}f_n^i(\tau')W(\alpha'p\tau')-e^{-in\tau}\int_{-\infty}^{\tau}d\tau'\;e^{in\tau'}f_n^i(\tau')W(\alpha'p\tau')\right)\;.
\ee
(To see that this is a solution, act on both sides by $\partial_\tau^2+n^2$\;.)  Then from the asymptotic solutions \eqref{flatsol}, \eqref{flatsol2}, and the Bogoliubov transformation \eqref{Bog}, we find
\be
B_n^i=\frac{p^2\alpha'^2}{2in}\int_{-\tau_0}^{\tau_0}d\tau\;e^{in\tau}f_n^i(\tau)W(\alpha'p\tau)\;.
\label{Beq}
\ee

In the region $U>T$, the mass squared and number operators are given by
\begin{subequations}
\begin{align}
M_>^2&=\frac{1}{\alpha'}\sum_{n=1}^\infty\left(\alpha_{n>}^{i\dagger}\alpha_{n>}^{i}+\tilde\alpha_{n>}^{i\dagger}\tilde\alpha_{n>}^{i}\right)+m_0^2\;,\\
N_>&=\sum_{n=1}^\infty \frac{1}{n}\left(\alpha_{n>}^{i\dagger}\alpha_{n>}^{i}+\tilde\alpha_{n>}^{i\dagger}\tilde\alpha_{n>}^{i}\right)\;,
\end{align}
\end{subequations}
where $m_0^2$ is the tachyon  mass squared.  The expectation values in the ingoing ground state $\ket{0_<}$ are
\begin{subequations}
\begin{align}
\langle M_>^2\rangle&=\frac{\bra{0_<}M_>^2\ket{0_<}}{\braket{0_<}{0_<}}=m_0^2+\frac{2}{\alpha'}\sum_{n=1}^\infty\sum_in|B_n^i|^2\;,\\
\langle N_>\rangle&=\frac{\bra{0_<}N_>\ket{0_<}}{\braket{0_<}{0_<}}=2\sum_{n=1}^\infty\sum_i|B_n^i|^2\;.
\end{align}
\label{MNeq}
\end{subequations}

Substituting $y=n\tau$ into \eqref{Beq} and using \eqref{ntau}, we find that
\be
B_n^i=\pm\frac{k}{2i}\int_{-n\tau_0}^{n\tau_0}dy\;\frac{e^{iy}f_1^i(y)}{y^2}\approx i\frac{z-1}{2z^2}\int_{-\infty}^{\infty}dy\;\frac{e^{iy}f_1^i(y)}{y^2}\;,
\ee
which is independent of $n$.  Using \eqref{x0sol}, one can show with a Fourier transform that the integral is finite.  Then for $z\neq1$, each mode is excited equally and the mode number and mass squared operators diverge.  The string excitations vanish when $z=1$, as expected, since in this case  the spacetime is $AdS_4$.\footnote{As we remarked earlier,  the plane wave is not a good approximation to the geometry for $z=1$.} The excitation is also suppressed as $z\rightarrow \infty$ since the geometry then approaches $AdS_2 \times \R^2$.

\end{section}

\begin{section}{Discussion}

We have shown that the singularity at the origin of the Lifshitz spacetimes is not removed by the known higher dimensional embeddings in supergravity and string theory.  We also studied the propagation of test strings in the Lifshitz geometry.  If $z > 1$, all string modes are turned on equally and the strings become infinitely excited if they attempt to cross the  singularity.  If $z < 1$, strings do not cross the singularity, but instead become infinitely large classically in a finite time $\tau$.  This case also violates the null energy condition\footnote{To see this, consider a radial null vector $\ell^\mu=(r^{-z},r,\vec0)$.  Then $T_{\mu\nu}\ell^\mu\ell^\nu=R_{\mu\nu}\ell^\mu\ell^\nu=3(z-1)$.} so this spacetime may not be physical.  If $z=1$, the metric is $AdS$ and the strings pass through the Poincar\'{e} horizon.  Although we did not study the propagation of test strings directly in the higher dimensional embeddings, they will presumably behave in a similar way.  

The fact that test strings become infinitely excited shows that the Lifshitz singularity is not just a singularity in the sense of general relativity, but is also a singularity in string theory.    A small nonzero temperature will hide this singularity behind a smooth horizon, but the tidal forces on infalling strings will still be large (just like the black holes in \cite{Horowitz:1997ed}). 
 This indicates an instability in the spacetime since the initial test string is like a perturbation which becomes large and backreacts on the metric.  These string perturbations do not respect the Lifshitz symmetries, so even starting at zero temperature the nonrenormalization theorem of \cite{Adams:2008zk} does not apply.  The endpoint of this instability and the final resolution of the singularity remain unresolved. However, one likely effect is a breakdown of the scaling symmetry in the deep infrared. This is because the corrections to the Lifshitz geometry will become important when the curvature reaches the string scale or the Planck scale. The introduction of these new length scales is likely to modify the original scaling symmetry.

\end{section}
%----- Acknowledgements -------
\vskip 1cm
\centerline{\bf Acknowledgements}
\vskip .5 cm
It is a pleasure to thank Koushik Balasubramanian, Sean Hartnoll, and Jorge Santos for helpful discussions.  This work was completed during the Holographic Duality and Condensed Matter Physics program at the Kavli Institute for Theoretical Physics. This work was supported in part by NSF grants PHY08-55415 and PHY05-51164.

%----- Bibliography ---------
\singlespacing


\begin{thebibliography}{99}

%\cite{Maldacena98}
\bibitem{Maldacena98}
J. M. Maldacena, ``The large N limit of superconformal field theories and supergravity,"
Adv. Theor. Math. Phys. \textbf{2}, 231 (1998)  [Int. J. Theor. Phys. \textbf{38},
1113 (1999)] [arXiv:hep-th/9711200].
	
%\cite{Gubser:1998bc}
\bibitem{Gubser:1998bc}
  S.~S.~Gubser, I.~R.~Klebanov and A.~M.~Polyakov,
  ``Gauge theory correlators from non-critical string theory,''
  Phys.\ Lett.\  B {\bf 428}, 105 (1998)
  [arXiv:hep-th/9802109].
  %%CITATION = PHLTA,B428,105;%%
  
  %\cite{Witten:1998qj}
\bibitem{Witten:1998qj}
  E.~Witten,
  ``Anti-de Sitter space and holography,''
  Adv.\ Theor.\ Math.\ Phys.\  {\bf 2}, 253 (1998)
  [arXiv:hep-th/9802150].
  %%CITATION = 00203,2,253;%%

%\cite{Kovtun:2004de}
\bibitem{Kovtun:2004de}
  P.~Kovtun, D.~T.~Son, A.~O.~Starinets,
  ``Viscosity in strongly interacting quantum field theories from black hole physics,''
  Phys.\ Rev.\ Lett.\  {\bf 94}, 111601 (2005).
  [hep-th/0405231].

  %\cite{Hartnoll:2009sz}
\bibitem{Hartnoll:2009sz}
  S.~A.~Hartnoll,
  ``Lectures on holographic methods for condensed matter physics,''
  Class.\ Quant.\ Grav.\  {\bf 26}, 224002 (2009).
  [arXiv:0903.3246 [hep-th]].
  
  %\cite{McGreevy:2009xe}
  \bibitem{McGreevy:2009xe}
  J.~McGreevy,
  ``Holographic duality with a view toward many-body physics,''
  arXiv:0909.0518 [hep-th].
  %%CITATION = ARXIV:0909.0518;%%

%\cite{Kachru:2008yh}
\bibitem{Kachru:2008yh}
  S.~Kachru, X.~Liu, M.~Mulligan,
  ``Gravity Duals of Lifshitz-like Fixed Points,''
  Phys.\ Rev.\  {\bf D78}, 106005 (2008).
  [arXiv:0808.1725 [hep-th]].
  
  %\cite{Taylor:2008tg}
\bibitem{Taylor:2008tg}
  M.~Taylor,
  ``Non-relativistic holography,''
  [arXiv:0812.0530 [hep-th]].
  
  %\cite{Gubser:2009cg}
\bibitem{Gubser:2009cg}
  S.~S.~Gubser, A.~Nellore,
  ``Ground states of holographic superconductors,''
  Phys.\ Rev.\  {\bf D80}, 105007 (2009).
  [arXiv:0908.1972 [hep-th]].
  
  %\cite{Hartnoll:2010gu}
\bibitem{Hartnoll:2010gu}
  S.~A.~Hartnoll, A.~Tavanfar,
  ``Electron stars for holographic metallic criticality,''
  Phys.\ Rev.\  {\bf D83}, 046003 (2011).
  [arXiv:1008.2828 [hep-th]].

  %\cite{Mann:2009yx}
\bibitem{Mann:2009yx}
  R.~B.~Mann,
  ``Lifshitz Topological Black Holes,''
  JHEP {\bf 0906}, 075 (2009).
  [arXiv:0905.1136 [hep-th]].

%\cite{Bertoldi:2009vn}
\bibitem{Bertoldi:2009vn}
  G.~Bertoldi, B.~A.~Burrington, A.~Peet,
  ``Black Holes in asymptotically Lifshitz spacetimes with arbitrary critical exponent,''
  Phys.\ Rev.\  {\bf D80}, 126003 (2009).
  [arXiv:0905.3183 [hep-th]].

  %\cite{Balasubramanian:2009rx}
\bibitem{Balasubramanian:2009rx}
  K.~Balasubramanian, J.~McGreevy,
  ``An Analytic Lifshitz black hole,''
  Phys.\ Rev.\  {\bf D80}, 104039 (2009).
  [arXiv:0909.0263 [hep-th]].

%\cite{Danielsson:2009gi}
\bibitem{Danielsson:2009gi}
  U.~H.~Danielsson, L.~Thorlacius,
  ``Black holes in asymptotically Lifshitz spacetime,''
  JHEP {\bf 0903}, 070 (2009).
  [arXiv:0812.5088 [hep-th]].
  
  %\cite{Hartnoll:2009ns}
\bibitem{Hartnoll:2009ns}
  S.~A.~Hartnoll, J.~Polchinski, E.~Silverstein, D.~Tong,
  ``Towards strange metallic holography,''
  JHEP {\bf 1004}, 120 (2010).
  [arXiv:0912.1061 [hep-th]].
  
    %\cite{Balasubramanian:2010uk}
\bibitem{Balasubramanian:2010uk}
  K.~Balasubramanian, K.~Narayan,
  ``Lifshitz spacetimes from AdS null and cosmological solutions,''
  JHEP {\bf 1008}, 014 (2010).
  [arXiv:1005.3291 [hep-th]].

%\cite{Donos:2010tu}
\bibitem{Donos:2010tu}
  A.~Donos, J.~P.~Gauntlett,
  ``Lifshitz Solutions of D=10 and D=11 supergravity,''
  JHEP {\bf 1012}, 002 (2010).
  [arXiv:1008.2062 [hep-th]].

  %\cite{Gregory:2010gx}
\bibitem{Gregory:2010gx}
  R.~Gregory, S.~L.~Parameswaran, G.~Tasinato, I.~Zavala,
  ``Lifshitz solutions in supergravity and string theory,''
  JHEP {\bf 1012}, 047 (2010).
  [arXiv:1009.3445 [hep-th]].
  
  %\cite{Copsey:2010ya}
\bibitem{Copsey:2010ya}
  K.~Copsey, R.~Mann,
  ``Pathologies in Asymptotically Lifshitz Spacetimes,''
  JHEP {\bf 1103}, 039 (2011).
  [arXiv:1011.3502 [hep-th]].
    
  %\cite{Adams:2008zk}
\bibitem{Adams:2008zk}
  A.~Adams, A.~Maloney, A.~Sinha, S.~E.~Vazquez,
  ``1/N Effects in Non-Relativistic Gauge-Gravity Duality,''
  JHEP {\bf 0903}, 097 (2009).
  [arXiv:0812.0166 [hep-th]].

  %\cite{Horowitz:1989bv,}
\bibitem{Horowitz:1989bv}
  G.~T.~Horowitz, A.~R.~Steif,
  ``Space-Time Singularities in String Theory,''
  Phys.\ Rev.\ Lett.\  {\bf 64}, 260 (1990);
  ``Strings In Strong Gravitational Fields,''
  Phys.\ Rev.\  {\bf D42}, 1950-1959 (1990).

%\cite{deVega:1990ke}
\bibitem{deVega:1990ke}
  H.~J.~de Vega, N.~G.~Sanchez,
  ``Strings falling into space-time singularities,''
  Phys.\ Rev.\  {\bf D45}, 2783-2793 (1992).
  
  %\cite{Cassani:2011sv}
\bibitem{Cassani:2011sv}
  D.~Cassani, A.~F.~Faedo,
  ``Constructing Lifshitz solutions from AdS,''
  JHEP {\bf 1105}, 013 (2011).
  [arXiv:1102.5344 [hep-th]].
  
  %\cite{Halmagyi:2011xh}
\bibitem{Halmagyi:2011xh}
  N.~Halmagyi, M.~Petrini, A.~Zaffaroni,
  ``Non-Relativistic Solutions of N=2 Gauged Supergravity,''
  JHEP {\bf 1108}, 041 (2011).
  [arXiv:1102.5740 [hep-th]].
  
  %\cite{Narayan:2011az}
\bibitem{Narayan:2011az}
  K.~Narayan,
  ``Lifshitz-like systems and AdS null deformations,''
  Phys.\ Rev.\  {\bf D84}, 086001 (2011).
  [arXiv:1103.1279 [hep-th]].
  
  %\cite{Chemissany:2011mb}
\bibitem{Chemissany:2011mb}
  W.~Chemissany, J.~Hartong,
  ``From D3-Branes to Lifshitz Space-Times,''
  Class.\ Quant.\ Grav.\  {\bf 28}, 195011 (2011).
  [arXiv:1105.0612 [hep-th]].
  
   %\cite{Gibbons:1994vm}
\bibitem{Gibbons:1994vm}
  G.~W.~Gibbons, G.~T.~Horowitz, P.~K.~Townsend,
  ``Higher dimensional resolution of dilatonic black hole singularities,''
  Class.\ Quant.\ Grav.\  {\bf 12}, 297-318 (1995).
  [hep-th/9410073].
  
  %\cite{Singh:2010zs}
\bibitem{Singh:2010zs}
  H.~Singh,
  ``Special limits and non-relativistic solutions,''
  JHEP {\bf 1012}, 061 (2010).
  [arXiv:1009.0651 [hep-th]].
  
  %\cite{Donos:2010ax}
\bibitem{Donos:2010ax}
  A.~Donos, J.~P.~Gauntlett, N.~Kim, O.~Varela,
  ``Wrapped M5-branes, consistent truncations and AdS/CMT,''
  JHEP {\bf 1012}, 003 (2010).
  [arXiv:1009.3805 [hep-th]].
  
  

%\cite{Horowitz:1997ed}
\bibitem{Horowitz:1997ed}
 G.~T.~Horowitz, S.~F.~Ross,
  ``Properties of naked black holes,''
  Phys.\ Rev.\  {\bf D57}, 1098-1107 (1998).
  [hep-th/9709050].
  

\end{thebibliography}
\end{document}